\documentstyle[12pt]{article}
\begin{document}

\begin{flushright}
NYU-TH/95-02\\[-0.2cm]
RAL-TR/95-021\\[-0.2cm]
June 1995
\end{flushright}

\begin{center}
{\bf{\LARGE Gauge Invariance and Unstable Particles}}\\[1.cm]
{\large Joannis Papavassiliou}$^a$
{\large and Apostolos Pilaftsis}$^b$\footnote[1]{E-mail address:
pilaftsis@v2.rl.ac.uk}\\[0.4cm]
$^a${\em Department of Physics, New York University, 4 Washington Place,
New York,\\
NY 10003, USA}\\[0.3cm]
$^b${\em Rutherford Appleton Laboratory, Chilton, Didcot, Oxon, OX11 0QX, UK}
\end{center}
\vskip0.7cm
\centerline {\bf ABSTRACT}

\noindent
A gauge-independent approach to resonant transition amplitudes with
nonconserved external currents is presented, which is implemented by
the pinch technique. The
analytic expressions derived with this method are $U(1)_{em}$ invariant,
independent of the choice of the gauge-fixing parameter,
and satisfy a number of required theoretical properties,
including unitarity.
Although special attention is paid to resonant scatterings involving
the $\gamma WW$ and $ZWW$ vertices in the minimal Standard Model, our
approach can be extended to the top quark or other unstable particles
appearing in renormalizable models of new physics.\\[0.4cm]
PACS nos.: 14.70.Fm, 11.15.Bt, 11.15.Ex

\newpage
Three decades after Veltman's pioneering work \cite{Velt},
the correct treatment of unstable
particles
in the context of renormalizable gauge field theories
is still an open question.
The interest in the problem resurfaced in recent years
\cite{WV},
mainly motivated by a plethora of phenomenological
applications linked to machines, such as the CERN Large Electron Positron
collider (LEP), the LEP2, planned to operate at centre of mass
system (c.m.s.) energy $s=200$ GeV, the TEVATRON at Fermilab,
and the CERN Large Hadron Collider (LHC).

Even though the need for a resummed propagator is evident
when dealing with unstable particles within the framework of the
$S$-matrix perturbation theory, its incorporation to
the amplitude of a resonant process is non-trivial.
When this incorporation is done naively, {\em e.g.}
by simply replacing the bare propagators of a
tree-level amplitude by resummed propagators,
one is often unable to satisfy basic field theoretical
requirements, such as the gauge-parameter independence of the
resulting $S$-matrix element, $U(1)_{em}$ symmetry,
high-energy unitarity, and the optical theorem.
This fact is perhaps not so surprising, since the
naive resummation of
the self-energy graphs takes into account higher order corrections, for
{\em only} certain parts of the tree-level amplitude.
Even though the amplitude possesses all the
desired properties,
this unequal treatment of its parts
distorts subtle cancellations,
resulting in numerous pathologies, which are artifacts of the
method used.
It is therefore important to devise a
self-consistent calculational scheme, which
{\em manifestly} preserves
the afore-mentioned field theoretical properties
that are {\it intrinsic} in every $S$-matrix element.

In this paper, we present a new gauge-independent (g.i.) approach
to resonant transition amplitudes implemented
by the pinch technique (PT)~\cite{PinchGeneral}.
The PT is an algorithm which systematically exploits all the
healthy properties of the $S$ matrix and hence has numerous applications
in electro-weak physics. Operationally, it amounts to a
rearrangement of the Feynman graphs contributing to
a g.i.\ amplitude so as to define {\em individually} g.i.\
propagator-, vertex-, and box-like structures.
In particular,  due to technical limitations,
one often attempts to cast the higher order corrections
to a given process in the form of its tree-level amplitude,
by essentially omitting box diagrams.
Since box graphs are however crucial for the final gauge independence
(among other things) of the $S$ matrix, their omission introduces
artificial gauge-dependences into the rest of the amplitude.
The way the PT circumvents such problems is by recognizing that
the parts of the boxes, which are instrumental for restoring the
final gauge invariance, are effectively propagator-like and
vertex-like, and thus can be naturally cast
in the form of the original
tree-level amplitude. The remaining box contributions are
a g.i.\ subset and can be consistently subtracted
out.
Even though the situation is conceptually and technically more subtle,
the underlying objective remains the
same when dealing with resonant amplitudes.
Again, one attempts to account for resonant effects
by ``dressing up" the tree-level amplitude; when
this is done without a concrete guiding principle, one ends up
with the type of pathologies mentioned above.

The crucial novelty we introduce here is that
the resummation of graphs must take place only
{\em after} the amplitude of interest has been cast
via the PT algorithm into
manifestly g.i.\ sub-amplitudes, with distinct
kinematic properties, order by order in perturbation theory.
The application of the PT remarkably remedies all the afore-mentioned
field-theoretical problems existing at present in the
literature. In particular,\\
(i) The analytic results obtained within our approach
are, by construction, {\em independent} of the
gauge-fixing parameter, in {\em every} gauge-fixing scheme
($R_{\xi}$ gauges, axial gauges, background field method, {\em etc.}).
In addition, by virtue of the tree-level Ward identities
satisfied by the PT Green's functions,
the $U(1)_{em}$ invariance can be enforced, without
introducing  residual gauge-dependent terms of higher orders.\\
(ii) The PT treats bosonic and fermionic contributions to the resummed
propagator of the $W$-, $Z$-boson, $t$ quark or other unstable particles,
{\em on equal footing}.
This feature is highly desirable for applications
to extensions of the SM at high energy colliders, such as
the LHC. For example, a heavy Higgs boson in the SM or
new gauge bosons, such as {\em e.g.} $Z'$, $W'$, $Z_R$, {\em etc.},
predicted in models beyond the SM, can have widths
predominantly originating from bosonic channels. In this way,
it becomes even more obvious that prescriptions
based on resumming {\em only} fermionic contributions as g.i.\ subsets of
graphs, are insufficient.\\
(iii) The use of an expansion of the resonant
matrix element in terms of a {\em constant} complex pole
produces unavoidably space-like threshold terms to all
orders, while non-resonant corrections remove such terms only up
to a given order.
These space-like terms, which explicitly violate unitarity,
manifest themselves when the c.m.s.\ energy of the process does
not coincide with the position of the resonant pole.
On the contrary, the PT circumvents these difficulties
by giving rise to an energy-dependent complex-pole
regulator.
For instance, possible unphysical absorptive parts originating from
channels below their production threshold have already been eliminated
by the corresponding kinematic $\theta$ functions.\\
(iv) Lastly, the amplitude obtained from our approach
exhibits a good high-energy unitarity  behaviour, as the c.m.s.\
energy $s\to \infty$. In fact, far away from the resonance,
the resonant amplitude tends to the usual PT amplitude, thus displaying
the correct high-energy unitarity limit of the entire tree-level
process.

We will now study some characteristic examples.
Within the PT framework, the transition
amplitude $T(s,t,m_i)$ of a $2\to 2$ process,
such as $e^-\bar{\nu}_e\to \mu^- \bar{\nu}_\mu$ with massive external
charged leptons, can be decomposed as
\begin{equation}
\label{TPT}
T(s,t,m_i)\ =\ \widehat{T}_1(s)\ +\ \widehat{T}_2(s,m_i)\ +\
\widehat{T}_3(s,t,m_i),
\end{equation}
where the piece $\widehat{T}_1$ contains three {\em individually}
g.i.\ quantities:
The $WW$ self-energy $\widehat{\Pi}^W_{\mu\nu}$, the $WG$ mixing
term $\widehat{\Theta}_{\mu}$, and the $GG$ self-energy
$\widehat{\Omega}$.
Similarly, $\widehat{T}_2(s,m_i)$ consists of two pairs of
g.i.\ vertices $W e^-\bar{\nu}_e$, $G e^-\bar{\nu}_e$
[${\widehat{\Gamma}}_{\mu}^{(1)}$ and ${\widehat{\Lambda}}^{(1)}$],
and $W\mu^-\bar{\nu}_\mu$ and $G\mu^-\bar{\nu}_\mu$
[${\widehat{\Gamma}}_{\mu}^{(2)}$ and ${\widehat{\Lambda}}^{(2)}$].
Most importantly, in addition to being g.i.,
the PT self-energies and vertices satisfy the following
{\em tree-like} Ward identities:
\begin{eqnarray}
q^\mu\widehat{\Pi}^W_{\mu\nu}\, -\, M_W\widehat{\Theta}_\nu &=& 0,\nonumber\\
q^\mu \widehat{\Theta}_\mu\, -\, M_W\widehat{\Omega} &=& 0, \nonumber\\
q^\mu \widehat{\Gamma}_{\mu}^{i}\, -\, M_W\widehat{\Lambda}^{i} &=& 0 ~(i=1,2).
\end{eqnarray}
These Ward identities are a direct consequence of the requirement
that $\widehat{T}_1$ and $\widehat{T}_2$ are fully $\xi$
independent~\cite{PinchGeneral}.
If we assume that the PT decomposition in Eq.~(\ref{TPT}) holds to any
order in perturbation theory
(the validity of this assumption will be discussed
extensively in Ref.~\cite{JPAP}),
and sum up contributions from all orders, we obtain for $\widehat{T}_1$
(suppressing contraction of Lorentz indices)
\begin{equation}
\widehat{T}_1 = \Gamma_0 U_W \Gamma_0 + \Gamma_0 U_W\widehat{\Pi}^WU_W \Gamma_0
+\Gamma_0 U_W \widehat{\Pi}^W\cdots \widehat{\Pi}^W U_W \Gamma_0
=\Gamma_0 \widehat{\Delta}_W \Gamma_0, \label{T1res}
\end{equation}
where $U_{W\mu\nu}(q)=t_{\mu\nu}(q)(q^2-M^2_W)^{-1}+\ell_{\mu\nu}(q)M^{-2}_W$
[$t_{\mu\nu}(q)=-g_{\mu\nu} +q_\mu q_\nu/q^2$ and
$\ell_{\mu\nu}(q)=q_\mu q_\nu/q^2$], and
\begin{equation}
\widehat{\Delta}_{W\mu\nu}(q)\ =\ \frac{t_{\mu\nu}(q)}{q^2 -
M_W^2 - \widehat{\Pi}^W_T (q^2)}\ -\ \frac{\ell_{\mu\nu}(q)}{
M_W^2-\widehat{\Pi}^W_L(q^2)}. \label{DeltaPT}
\end{equation}
In Eq.~(\ref{DeltaPT}), we have decomposed $\widehat{\Pi}^W_{\mu\nu}=t_{\mu\nu}
\widehat{\Pi}^W_T +\ell_{\mu\nu}\widehat{\Pi}^W_L$.

Next we apply our formalism to the
process $\gamma e^- \to \mu^-\bar{\nu}_\mu \nu_e$,
in which two $W$ gauge bosons are involved. This process is of
potential interest at the LEP2. We concentrate on the
part of the amplitude
($\widehat{T}_{1\mu}$) involving the $\gamma WW$ vertex,
as given in Fig.~1. As discussed above, the PT
method reorders the Feynman graphs
into manifestly g.i.\ subsets.
Resumming the PT self-energies
one obtains the following resonant transition amplitude:
\begin{equation}
\label{Twres}
\widehat{T}_{1\mu}\ =\
\Gamma_0 \widehat{\Delta}_W (\Gamma^{\gamma W^- W^+}_{0\, \mu}+
\widehat{\Gamma}^{\gamma W^- W^+}_\mu) \widehat{\Delta}_W \Gamma_0\ +\
\Gamma_0 S^{(e)}_0 \Gamma^\gamma_{0\, \mu}\widehat{\Delta}_W \Gamma_0\
+\ \Gamma_0 \widehat{\Delta}_W \Gamma^\gamma_{0\, \mu} S^{(\mu )}_0 \Gamma_0,
\end{equation}
where $S^{(f)}$ is the free $f$-fermion propagator and $\Gamma^\gamma_{0\mu}$
is the tree $\gamma ff$ coupling. In Eq.~(\ref{Twres}),
contraction over all Lorentz indices except of the photonic one
is implied.
Since the action of the photonic momentum ($q$) on the tree-level and
one-loop PT $\gamma WW$ vertices gives:
$\frac{1}{e}\, q^\mu \Gamma^{\gamma W^-W^+}_{0\, \mu\nu\lambda}\ =\
U^{-1}_{W\nu\lambda}(p_+) - U^{-1}_{W\nu\lambda}(p_-)$ and
$\frac{1}{e}\, q^\mu \widehat{\Gamma}^{\gamma W^-W^+}_{ \mu\nu\lambda}\ =\
\widehat{\Pi}^W_{\nu\lambda}(p_-) - \widehat{\Pi}^W_{\nu\lambda}(p_+)$,
respectively,
the $U(1)_{em}$ gauge invariance of this resonant process is restored,
{\em i.e.}\ $q^\mu \widehat{T}_{1\mu}=0$.
To any loop order, $U(1)_{em}$ and $R_\xi$ invariance are warranted
by virtue of the tree-type Ward identities that the PT vertex $\gamma WW$
satisfy (all momenta flow into the vertex, {\em i.e.}, $q+p_-+p_+=0$):
\begin{eqnarray}
\frac{1}{e}q^{\mu}\widehat{\Gamma}_{\mu\nu\lambda}^{\gamma W^- W^+}&=&
\widehat{\Pi}^W_{\nu\lambda} (p_-)-\widehat{\Pi}^W_{\nu\lambda} (p_+),
\nonumber\\
\frac{1}{e}[p^\nu_-\widehat{\Gamma}_{\mu\nu\lambda}^{\gamma W^- W^+}
-M_W \widehat{\Gamma}_{\mu\lambda}^{\gamma G^- W^+}]
&= &\widehat{\Pi}_{\mu\lambda}^{W}(p_+)-\widehat{\Pi}_{\mu\lambda}^{\gamma}(q)
- \frac{c_w}{s_w}\widehat{\Pi}_{\mu\lambda}^{\gamma Z}(q), \nonumber\\
\frac{1}{e}[p^\lambda_+\widehat{\Gamma}_{\mu\nu\lambda}^{\gamma W^- W^+}
+ M_W \widehat{\Gamma}_{\mu\nu}^{\gamma W^- G^+}]
&= &-\widehat{\Pi}_{\mu\nu}^{W}(p_-)+\widehat{\Pi}_{\mu\nu}^{\gamma}(q)
+\frac{c_w}{s_w}\widehat{\Pi}_{\mu\nu}^{\gamma Z}(q). \label{PTWWgamma}
\end{eqnarray}

Of particular interest in testing electroweak theory at TEVATRON is
the process $QQ'\to e^-\bar{\nu}_e \mu^-\mu^+$. In addition to
the $\gamma WW$
vertex, the $ZWW$ coupling is now important, especially when
the invariant-mass cut $m(\mu^-\mu^+)\simeq M_Z$ is imposed.
Noticing that the PT self-energy of the photon and $Z\gamma$
mixing are transverse [$q^\mu\widehat{\Pi}^{\gamma Z}_{\mu\lambda}(q)
= q^\mu\widehat{\Pi}^\gamma_{\mu\lambda}(q) = 0$], we find that the
part of $\widehat{T}_1$, in which all tree photonic interactions
are absent as shown in Fig.~2, is individually g.i., having
the form
\begin{eqnarray}
\widehat{T}^Z_1 &=& \Gamma_0 \widehat{\Delta}_W(p_-) (\Gamma^{ZW^- W^+}_0
+\widehat{\Gamma}^{Z W^- W^+})\widehat{\Delta}_Z(q)\Gamma^Z_0
\widehat{\Delta}_W(p_+) \Gamma_0\nonumber\\
&&+\
\Gamma_0 S^{{}^{(Q)}}_0 \Gamma^Z_0\widehat{\Delta}_Z(q)\Gamma^Z_0
\widehat{\Delta}_W(p_+) \Gamma_0\
+\ \Gamma^Z_0 S^{{}^{(Q')}}_0 \Gamma_0 \widehat{\Delta}_Z(q)\Gamma^Z_0
\widehat{\Delta}_W(p_+) \Gamma_0\nonumber\\
&&+\ \Gamma_0 \widehat{\Delta}_W(p_-) \Gamma^Z_0
\widehat{\Delta}_Z(q)\Gamma^Z_0 S^{(e)}_0 \Gamma_0 \
+\ \Gamma_0 \widehat{\Delta}_W(p_-) \Gamma^Z_0 \widehat{\Delta}_Z(q)\Gamma_0
S^{(\nu_e)}_0\Gamma^Z_0\nonumber\\
&&+\ \Gamma_0\widehat{\Delta}_W(p_-) \Gamma_0 S^{(\nu_\mu )}_0 \Gamma_0
\widehat{\Delta}_W(p_+)\Gamma_0,
\label{hatTZ1}
\end{eqnarray}
where $\Gamma^Z_0$ stands for the tree $Zff$ coupling.
The PT Ward identities maintaining the gauge invariance of this process have
been derived in~\cite{JPKP}.

We continue with some important technical remarks.
We first focus on issues of resummation, and argue that the g.i.\
PT self-energy may be resummed, exactly as one carries out
the Dyson summation of the
conventional self-energy.
The crucial point is that,
even though contributions from vertices and boxes are instrumental
for the definition of the PT self-energies,
their resummation
does {\em not} require a corresponding resummation
of vertex
or box parts. In order to construct g.i.\ chains of
self-energy bubbles, one can borrow pinch contributions from
existing graphs, which are however not sufficient to convert
each $\Pi_{\mu\nu}$ in the chain into the corresponding
$\widehat{\Pi}_{\mu\nu}$. If we add the missing pieces by hand, and
subsequently subtract them, we notice that: {\bf (i)} The regular
string has been converted into a g.i.\ string, with
$\Pi_{\mu\nu}\to \widehat{\Pi}_{\mu\nu}$, and
{\bf (ii)} The left-overs, due to the characteristic
presence of the inverse bare propagator,
$(U^{\mu\nu})^{-1}$, are
effectively one-particle {\it irreducible}.
In fact, the left-over terms have the same structure as the
one-particle irreducible self-energy graphs,
and together with
the genuine vertex
($V^{P}$) and box pinch contributions ($B^{P}$) will
convert the conventional self-energy into the
g.i. PT self-energy.
This procedure is generalizable to an arbitrary order.
So, the transverse propagator-like pinch
contributions in the Feynman gauge,
to a given order $n$ in perturbation theory,
have the general form
\begin{equation}
\Pi^{P}_{n}(q^{2})= (q^{2}-m_{0}^{2})V_{n}^{P}(q^{2})+
{(q^{2}-m_{0}^{2})}^{2}B_{n}^{P}(q^{2})+R_{n}^{P}(q^{2})\, ,
\label{GFPT}
\end{equation}
where $R_{n}^{P}$ are the residual pieces of order $n$.
For $n=2$, for example, it is easy to check that the string
$(\frac{1}{q^{2}-m_{0}^{2}}){\Pi}
(\frac{1}{q^{2}-m_{0}^{2}}){\Pi}(\frac{1}{q^{2}-m_{0}^{2}})$,
together with
existing pinch pieces from graphs containing vertices, needs
an additional amount $-R_{2}^{P}$, given by
\begin{equation}
-R_{2}^{P}(q^{2})\ =\
\Pi V^{P}_1+ \frac{3}{4}(q^{2}-m_{0}^{2})
V^{P}_1V^{P}_1 \,~,
\label{R2}
\end{equation}
in order to be converted into the g.i.\ string
$(\frac{1}{q^{2}-m_{0}^{2}})\widehat{\Pi}
(\frac{1}{q^{2}-m_{0}^{2}})\widehat{\Pi}(\frac{1}{q^{2}-m_{0}^{2}})$.
However, $R_{2}^{P}$ will be absorbed by the one-particle irreducible
two-loop self-energy shown in Fig.~3. In general,
the $R_{n}^{P}$ terms consist of products of lower order
conventional self-energies $\Pi_{k}(q^{2})$,
and lower order pinch contributions $V^{P}_{\ell}$ and (or)
$B^{P}_{\ell}$ , with $k+\ell=n$~\cite{JPAP}.

Another issue is whether the g.i.\ PT complex pole is identical to
the g.i.\ physical pole of the amplitude. Here we concentrate on the
case of a stable particle, and demonstrate how its mass does not
get shifted by the PT.
The masses $m$ and $\widehat{m}$ are respectively defined as the solution
of the equations: $m^{2}=m_{0}^{2}+\Pi(m^{2})$ and
${\widehat{m}}^{2}=m_{0}^{2}+\widehat{\Pi}({\widehat{m}}^{2})$.
In perturbation theory, $m^{2}=m_{0}^{2}+\sum_{1}^{\infty}g^{2n}C_{n}$
and ${\widehat{m}}^{2}=m_{0}^{2}+\sum_{1}^{\infty}g^{2n}{\widehat{C}}_{n}$,
and one has hence to show that $C_n-\widehat{C}_n=O(g^{2n+1})$.
To zeroth order $m^{2}={\widehat{m}}^{2}=m_{0}^{2}$. Similarly,
from Eq.~(\ref{GFPT}), using the fact that $B_{1}^{P}=0$
(in the Feynman gauge),
and $R_{1}^{P}$=0 (in any gauge), we have that $C_{1}={\widehat{C}}_{1}$,
because the pinch contribution $(q^{2}-m_{0}^{2})V_{1}^{P}$ is of
$O(g^{4})$.
The non-trivial step in generalizing this proof to higher orders
is to observe that
not all pinch contributions of Eq.~(\ref{GFPT}) contribute terms of higher
order. To be precise, the terms of $R^P_n$ which do not have the
characteristic factor $(q^{2}-m_{0}^{2})$ in front
are {\it not} of higher order, and are instrumental for our proof.
We will illustrate this point at the two-loop
order.
The second order $m^{2}$ and ${\widehat{m}}^{2}$ are given by:
\begin{eqnarray}
m^{2}&=&m_{0}^{2}+\Pi_{1}(m^{2})+\Pi_{2}(m^{2})\nonumber\\
{\widehat{m}}^{2} &=& m_{0}^{2}+\Pi_{1}({\widehat{m}}^{2})+
\Pi_{2}({\widehat{m}}^{2})+ \Pi_{1}^{P}+ \Pi_{2}^{P}\nonumber
\end{eqnarray}
where the subscripts 1 and 2 denote loop order, and
\begin{eqnarray}
\Pi_{1}^{P}({\widehat{m}}^{2})+\Pi_{2}^{P}({\widehat{m}}^{2})
&=& ({\widehat{m}}^{2}-m_{0}^{2})
[ V_{1}^{P}({\widehat{m}}^{2})+V_{2}^{P}({\widehat{m}}^{2}) ] +
{({\widehat{m}}^{2}-m_{0}^{2})}^{2}
[ B_{1}^{P}({\widehat{m}}^{2})+B_{2}^{P}({\widehat{m}}^{2})]\nonumber\\
&&+ R_{2}^{P}({\widehat{m}}^{2})\, .
\label{SP6}
\end{eqnarray}
It is not difficult to show that
$\Pi_{1}^{P}({\widehat{m}}^{2})+\Pi_{2}^{P}({\widehat{m}}^{2})=O(g^{6})$.
Substituting~ ${\widehat{m}}^{2}-m_{0}^{2}=\Pi_{1}({\widehat{m}}^{2})+
O(g^{4})$
into Eq.~(\ref{SP6}), and neglecting terms of
$O(g^{6})$ or higher, we find
\begin{displaymath}
\Pi_1^{P}({\widehat{m}}^{2})+\Pi_2^{P}({\widehat{m}}^{2})=
R_{2}^{P}({\widehat{m}}^{2})+
\Pi_{1}({\widehat{m}}^{2})V^P_{1}({\widehat{m}}^{2})+
O(g^{6})=0+O(g^{6}),
\end{displaymath}
where we have also used Eq.~(\ref{R2}) at
$q^{2}={\widehat{m}}^{2}$.
The generalization of the proof to an arbitrary
order $n$ in perturbation theory proceeds by induction
and will be given in Ref.~\cite{JPAP}, together with
the case of an unstable particle---both mass and width remain unshifted.

Another point, important for unitarity, is whether
the PT self-energy
contains any unphysical absorptive parts. In particular,
the propagator-like part $\widehat{T}_1$ of a
reaction should contain imaginary parts associated with
physical Landau singularities only, whereas the unphysical
poles related to Goldstone bosons and ghosts must vanish in the loop.
Explicit calculations (see Ref.~\cite{JPAP})
show that, indeed, our
g.i.\ procedure does not introduce any
fixed unphysical poles.
Here we offer only a qualitative
argument in that vein, namely that
the PT results
may be obtained equally well if one  works {\it directly} in the
unitary gauge, where only physical Landau poles are present.

Although our discussion has been restricted to the $W$ and $Z$
gauge bosons, our considerations are also valid for the
heavy top quark, and will provide a self-consistent framework
for investigating the CP properties of the
$t$ quark at LHC. Moreover,
our analytic g.i.\ approach can be straightforwardly extended
to analyze possible new-physics phenomena
induced by non-SM gauge bosons, such as the bosons $W_R$, $Z'$, {\em etc.},
predicted in $SO(10)$ or $E_6$ unified models~\cite{GUT}.
Since our analytic method treats bosonic
and fermionic contributions equally, it
can provide a consistent framework for the study of
the resonant dynamics of
a heavy Higgs boson and of a strong Higgs sector at the LHC.

\noindent
{\bf Acknowledgements.} The authors gratefully acknowledge discussions
with J.~M.~Cornwall,  K.~Philippides, A.~Sirlin,
R.~Stuart, and D.~Zeppenfeld.

\newpage

\newpage

\centerline{\bf\Large Figure Captions }
\vspace{-0.2cm}
\newcounter{fig}
\begin{list}{\bf\rm Fig. \arabic{fig}: }{\usecounter{fig}
\labelwidth1.6cm \leftmargin2.5cm \labelsep0.4cm \itemsep0ex plus0.2ex }

\item The process $e^-\gamma\to \mu^-\bar{\nu}_\mu\nu_e$ in
our PT approach.

\item The reaction $QQ'\to \mu^+\mu^-e^-\bar{\nu}_e$, where
photonic and $Z\gamma$-mixing graphs are not shown.

\item Typical two-loop self-energy graphs (a)--(d), and
some of the residual pinch contributions (e)--(h) contained
in $R_2^P$.

\end{list}

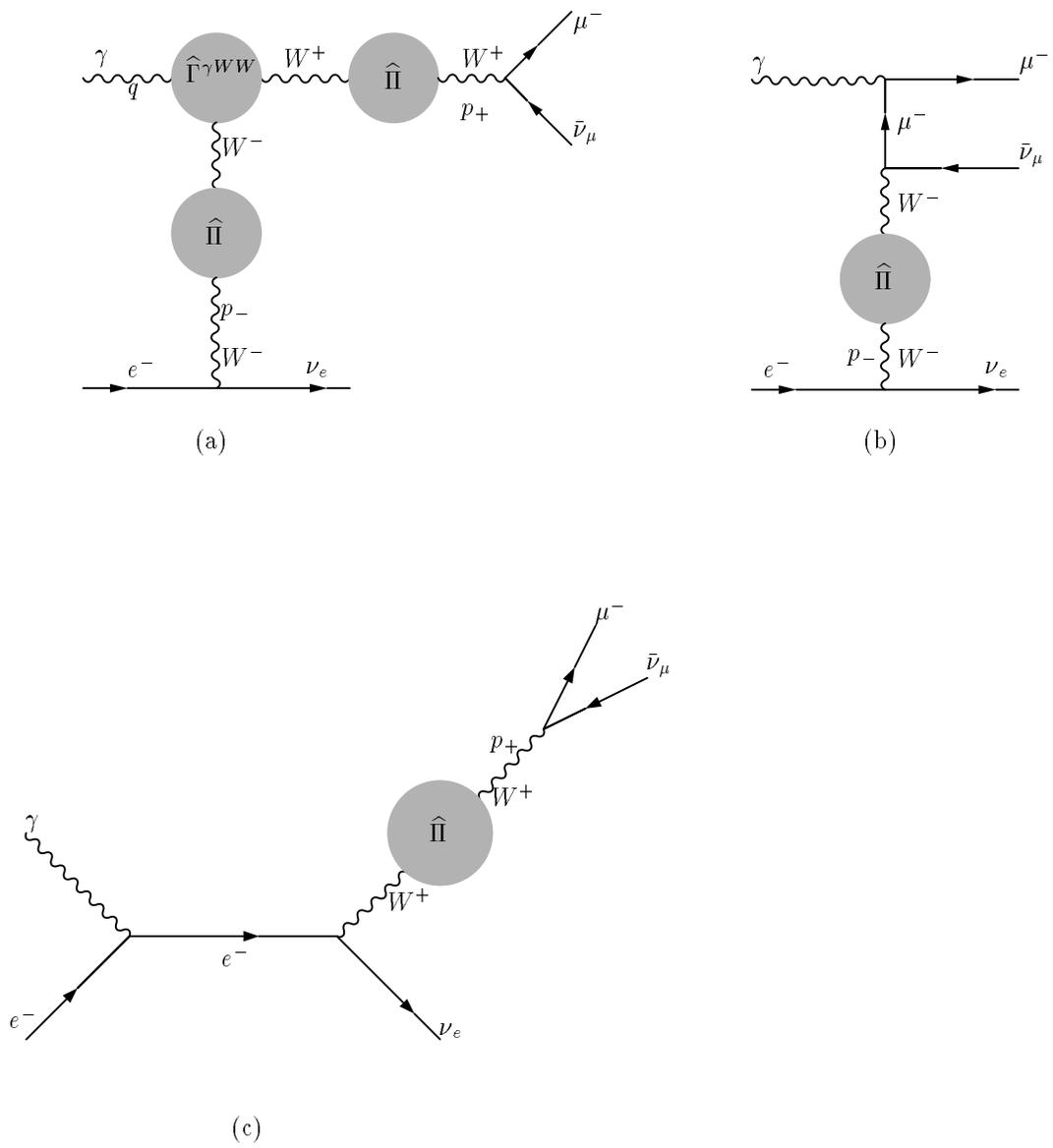
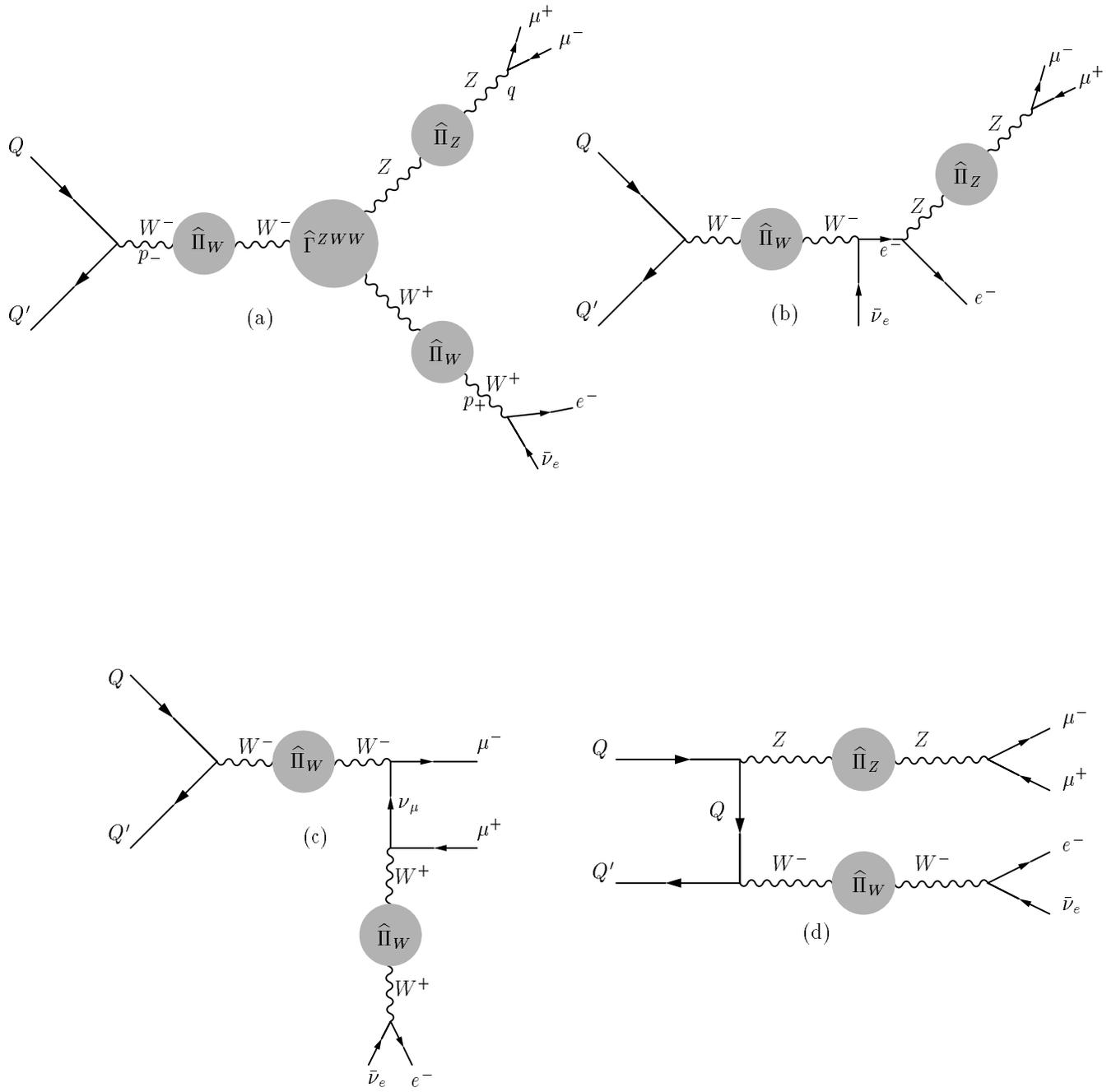
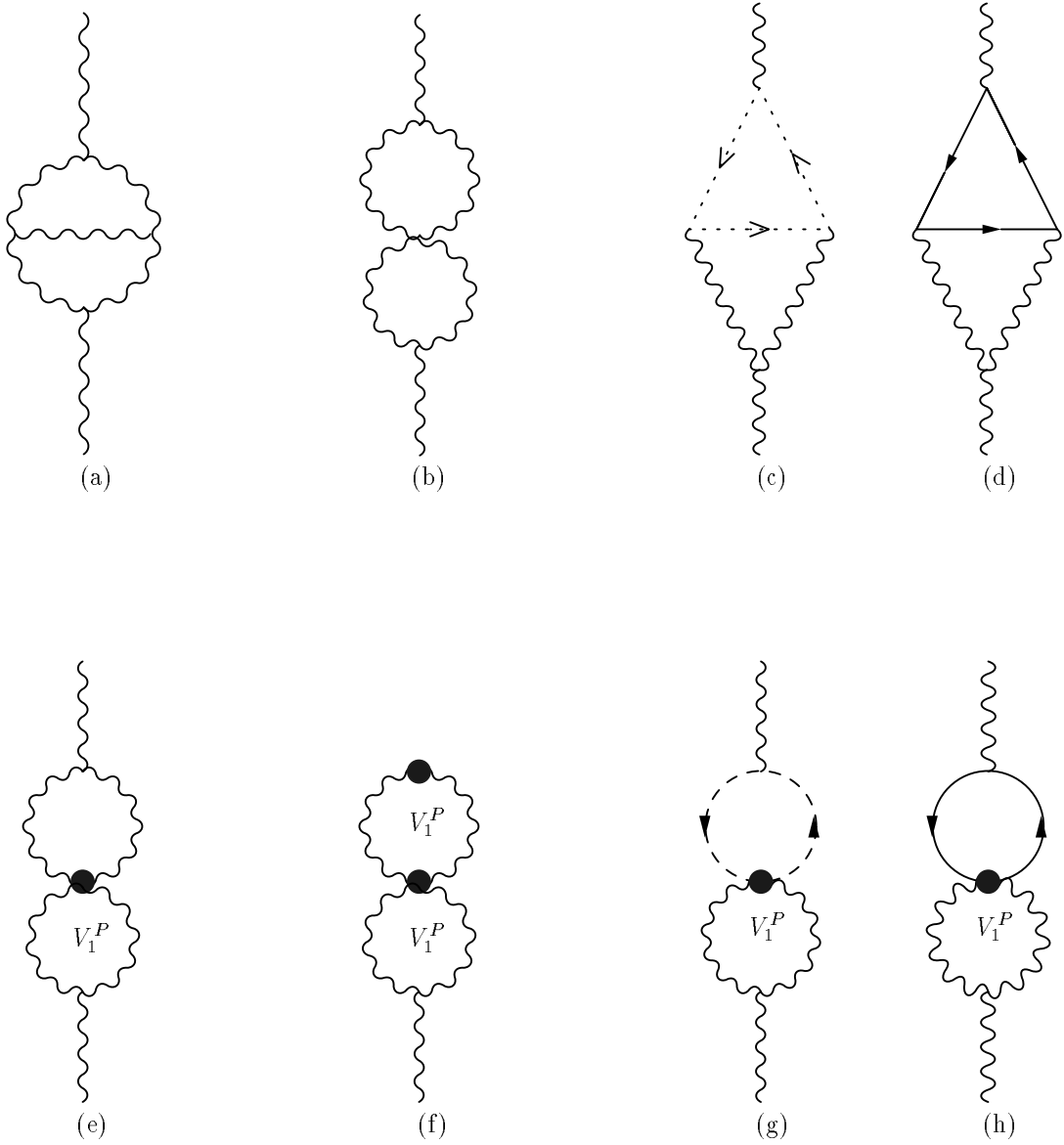
\end{document}